\documentclass[twocolumn,printer]{aa}

\usepackage{graphicx}
\begin{document}

\title{High-density QCD pairing in compact star structure}

\author{G. Lugones \inst{1}   \and  J. E. Horvath \inst{1}  }

\institute{Instituto de Astronomia, Geof\'{\i}sica e Ci\^encias Atmosf\'ericas,
Universidade de S\~ao Paulo\\
Rua do Mat\~ao 1226, (05508-900) S\~ao Paulo SP, Brazil\\
\email{glugones@astro.iag.usp.br ; foton@astro.iag.usp.br}}

\date{Received 2 December 2002 / Accepted 14 February 2003}

\abstract{Strange quark matter in a color flavor locked (CFL)
state can be the true ground state of hadronic matter for a much
wider range of the parameters of the model (the gap of the QCD
Cooper pairs $\Delta$, the strange quark mass $m_s$ and the Bag
Constant $B$) than the state without any pairing. We review the
equation of state (EOS) of CFL strange matter and study the
structure of stellar objects made up of this phase, highlighting
the novel features of the latter. Although the effects of pairing
on the equation of state are thought to be small, we find that CFL
stars may be in fact much more compact than strange stars (SS).
This feature may be relevant in view of some recent observations
claiming the existence of exotic and/or deconfined phases in some
nearby neutron stars (NS).
   }
\maketitle

\section{Introduction}

The study of a hypothetical stability of strange quark matter
(SQM), put forward in Witten's (\cite{Witten}) seminal paper and a
few important precursors (Bodmer \cite{Bodmer}; Terazawa
\cite{Terazawa}; Chin and Kerman \cite{Chin}) has entered its
third decade. Nothing less  than the nature of the true ground
state of hadronic matter is being questioned and in fact, within
simple models, the hypothesis of a stable form of cold catalyzed
plasma was shown to be tenable. Following the pioneering works, a
general calculation of strange matter by Farhi and Jaffe
(\cite{farhijaffe}) in the framework of the MIT Bag model of
confinement (Degrand et al. \cite{Degrand}; Cleymans, Suhonen and
Gavai \cite{Cleymans}) identified the so-called "windows of
stability", or regions in the plane $m_{s}-B$ inside which the
stability of SQM can be realized. Other models of confinement have
been worked out to find a quite ample range of conditions for SQM
to be absolutely bound (Zhang and Ru-Ken Su \cite{massdep} and
references therein; Hanauske et al.  \cite{NJL}). However, there
is a consensus that the issue of the availability of a $\sim 1 \,
\% $ binding energy difference for SQM to be bound is ultimately
an experimental matter.

While the search for SQM in laboratory and astrophysical
environments beyond their present limits continues, important
theoretical developments have taken place. The most sound is
probably the revival of interest in pairing interactions in dense
matter, a subject already addressed in the early 1980s (Bailin and
Love \cite{BL}) and revived a few years ago with new calculations
of the pairing energy and related physics. It is now generally
agreed (Alford, Rajagopal and Wilczek \cite{ARW}; Rapp et al.
\cite{rap}; Rajagopal and Wilczeck \cite{rajagopal2}) that (at
least for very high densities) the color-flavor locked (CFL)
state, with equal numbers of $u,d$ and $s$ quarks is likely to be
the ground state of strong interactions, even if the quark masses
are unequal (Alford, Berges and Rajagopal \cite{uneq}; Sch\"afer
and Wilczek \cite{Schafer}). The equal number of flavors is
enforced by the symmetry of the state, and thus electrons are
absent because the mixture is automatically neutral (Rajagopal and
Wilczeck \cite{rajagopal}; { Steiner, Reddy and Prakash 2002}).

Long awaited by theoretical physicists, the high-performance of
the space X-ray missions Chandra and XMM (Weisskopf
\cite{Chandra}; Becker and Aschenbach \cite{XMM}) enabled
unprecedented studies of imaging and spectra of selected neutron
stars with the aim of determining the masses and radii, perhaps
the most simple forms of (indirectly) investigating the nature of
high density matter. Adopting General Relativity as a framework, a
comparison of the static models generated by integration of the
Tolman-Oppenheimer-Volkof equations with observed data is expected
to give information about the equation of state, and possibly
other effects such as rotation, magnetic atmospheres and so on. A
wonderful example of the pre-space  determinations is the mass of
the binary pulsar PSR 1913+16, accurate to several decimal places
(van Kerkwijk \cite{vankerkwijk}). Other methods based on
combinations of spectroscopic and photometric techniques have been
recently proposed. At least one X-ray source is significantly
above the centroid of the binary distribution $\sim \, 1.4 \,
M_{\odot}$; namely Vela X-1 for which a value of
$1.87^{+0.23}_{-0.17} \, M_{\odot}$ has been obtained (van
Kerkwijk \cite{vankerkwijk}). Recently, not only the masses have
been determined, but also indications of the radii became
available, suggesting a very compact structure. For instance,
claims of high compactness have been made from the analysis of the
binary Her X-1 (Li, Lai and Wang \cite{li}; Dey et al.
\cite{Dey98}) with $M = 0.98 \pm 0.12 M_{\odot}$ and $R = 6.7 \pm
1.2$ km, and of the isolated nearby RX J185635-3754 (Pons et al.
\cite{Pons}) with $M \approx 0.9 M_{\odot}$ and $R \approx 6$ km.
In both cases, the results have been revisited and challenged by
other groups (Reynolds et al. \cite{Rey}; Kaplan, van Kerkwijk and
Anderson \cite{Kaplan}) who in turn found figures around the
expected for conventional neutron star models. This stresses the
cautionary remarks made by several researchers about the
high-compactness objects and guarantees further studies, already
undertaken in most cases. Li et al. (\cite{Li99b}) have also added
the source 4U 1728-34 to the candidate list, showing that
conventional accretion models indicate a very compact source in
the mass-radius plane. As always, the actual distance to the
source is a matter of concern. Needless to say, these results have
yet to be confirmed carefully. Nonetheless, it is worthwhile to
consider the possibility that at least some compact stars are
extremely compact, or in other words, that their radii are $\sim
30-40 \%$ less than the "canonical" $10 \, km$ favored by neutron
matter models for $M \, \sim 1 \, M_{\odot}$.

If actually present in these sources, compactness would be
extremely difficult (perhaps impossible) to model using underlying
equations of state based on ordinary hadrons alone, and a natural
alternative would be to consider deconfined matter (or other
exotic components, like kaon matter or hyperons). Stars suspected
to be made of deconfined matter also include the X-ray bursters
GRO J1744-28 (Cheng et al. \cite{Cheng98}) and SAX J1808.4-3658
(Li at al. \cite{Li99a}). This increasing availability of data on
compact neutron star structure coupled with the new developments
on the QCD ground sate leads to investigate the properties of
compact stars in the light of the CFL state \footnote{ { Alford
and Reddy (2002) have also independently analyzed  the role of CFL
quark matter in compact stars in a very recent work. Their work is
rather general and focused on stars with a normal matter envelope,
while our work deals with {\it absolutely stable} CFL stars only.
However, it is worth noting that Alford and Reddy (2002) also
include some results on pure CFL stars in their Figs. 5 and 6,
which are in quantitative agreement with ours (see Section 3).}}.

\section{Color flavor locked strange matter}

\subsection{The equation of state}

The very complex structure of the QCD phase diagram emerging from
detailed calculations by several groups is still being explored.
Therefore, only schematic models are available to explore stellar
structure questions. It is, however, widely agreed that if the
quark mass $m_{s}$ is small enough (possibly the actual case in
nature), the CFL state would be the minimum energy configuration
at high densities. To proceed we need the thermodynamical
potential $\Omega_{CFL}$ to derive the relevant quantities.
Fortunately, this potential can be found quite simply (Alford et
al. \cite{alfordcfl}) to the lowest order in the gap $\Delta^2$.
The procedure is to start with the $\Omega_{free}$ of a fictional
state of unpaired quark matter in which all quarks which are
"going to pair" have a common Fermi momentum $\nu$, with $\nu$
chosen to minimize $\Omega_{free}$. The binding energy of the
diquark condensate is included by subtracting the condensation
term $3 \Delta^2 \mu^2 / \pi^2$ while the vacuum energy is
introduced by means of the phenomenological bag constant $B$
allowing the mixture to confine. In this work we assume that the
CFL state has been reached by deconfined matter, although the
precise way in which this is done (involving weak decays, etc)
merits a closer look. The thermodynamic potential $\Omega_{CFL}$
of this model reads (Alford et al. \cite{alfordcfl})

\begin{eqnarray}
\nonumber \Omega_{CFL}& =& \Omega_{free} - \frac{3}{\pi^2} \Delta^2 \mu^2 + B \\
\nonumber &=& \sum_{i=u,d,s} \frac{1}{4 \pi^2} \bigg[ \mu_i \nu (\mu_i^2 - \frac{5}{2} m_i^2) +
\frac{3}{2} m_i^4 \log\bigg( \frac{\mu_i + \nu}{m_i} \bigg) \bigg] \\
& & - \frac{3}{\pi^2} \Delta^2 \mu^2 + B,
\end{eqnarray}

\noindent where $3 \mu = \mu_u + \mu_d + \mu_s$, $\mu_i$  being
the chemical potential of the i-species. The common Fermi momentum
$ \nu = ( \mu_i^2 - m_i^2)^{1/2}$ is given by

\begin{equation}
\nu = 2 \mu - \bigg( \mu^2 + \frac{m_s^2}{3} \bigg)^{1/2} \label{nu}.
\end{equation}

\noindent The thermodynamic quantities, such as the pressure P,
the baryon number density $n_{B}$, the particle number densities
$n_i$, and the energy density $\varepsilon$ can be easily derived
at $T=0$ and read (Lugones and Horvath \cite{cflprd})

\begin{equation}
P = - \Omega_{CFL} ,
\label{P}
\end{equation}

\begin{equation}
n_B = n_u = n_d = n_s = \frac{( \nu^3 + 2 \Delta^2 \mu ) }{\pi^2} ,
\label{nb}
\end{equation}

\begin{equation}
\varepsilon = \sum_i \mu_i n_i + \Omega_{CFL} = 3 \mu n_B - P .
\label{E}
\end{equation}

In this approach we shall treat the values of B, $m_s$ and
$\Delta$ (possibly as high as $\sim 100 \, MeV$) as free constant
parameters. The full dependence of $\Omega_{CFL}$ on $m_{s}$ has
been known for years and certainly complicates the evaluation of
the equation of state, which must be treated numerically. However,
we have recently worked out an approximation to the order
$m_{s}^{2}$ which has the main advantage of  keeping the equation
of state very simple, yet useful for most calculations, while at
the same time highlights the effect of each parameter of the model
(see Lugones and Horvath \cite{cflprd} for details).

The stiffness of the EOS is relevant for stellar models since it
has a direct impact on the compactness  of the stellar
configurations. As discussed by Lugones and Horvath
(\cite{cflprd}), whenever $\Delta$ is higher than $m_{s}/2$, the
EOS is stiffer than the unpaired SQM (that is, produces more
pressure for a given energy density). Since the actual value of
$\Delta$ is not well known, we may expect either case, a stiffer
or a softer EOS (for a given $B$).

\begin{figure}
\centering
\includegraphics[angle=-90,width=8cm,clip]{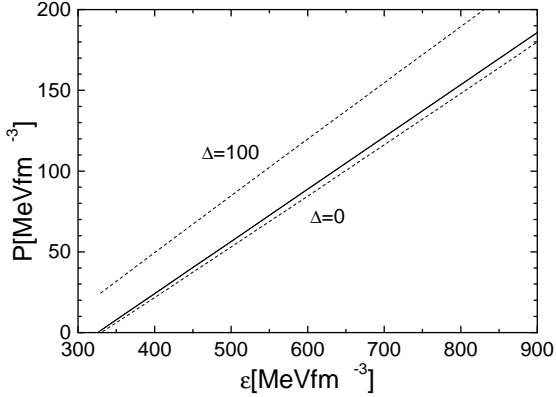}
\caption{The dashed lines show  the equation of state for CFL
strange matter for  B=75 MeV fm$^{-3}$,  $m_s$ = 150 MeV, and two
different values of the gap $\Delta$ as indicated in the figure.
The solid line corresponds to SQM without pairing. Note the change
of stiffness according to the value of $\Delta$, as discussed in
the text.}
\end{figure}

\begin{figure}
\centering
\includegraphics[angle=-90,width=8cm,clip]{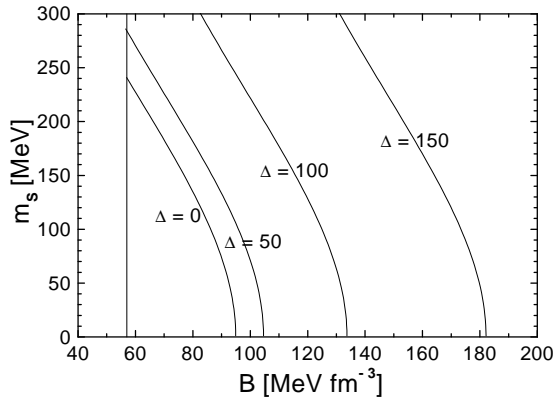}
\caption{The stability windows  for CFL strange matter. If the
strange quark mass $m_{s}$ and the bag constant $B$ lie inside the
bounded region  the CFL state is absolutely stable. Each window
corresponds to a given value of the gap $\Delta$ as indicated by
the label. The vertical solid line is the limit imposed by
requiring instability of two-flavor quark matter. Note the
enlargement of the window with increasing $\Delta$.}
\end{figure}

\subsection{Stability of the CFL phase}

Self-bound matter, in which strong forces produce a zero-pressure
point at finite density, is potentially interesting for
astrophysics because it produces stellar sequences having $M \,
\propto R^{3}$ in the Newtonian limit. For a given EOS the energy
per baryon $\varepsilon / n_B$ of the deconfined phase (at $P=0$
and $T=0$) must be lower than the neutron mass $m_n$ if matter is
to be absolutely stable. Another condition to be considered
results from the empirically known stability of normal nuclear
matter against deconfinement at  $P=0$ and $T=0$ (Farhi and Jaffe
\cite{farhijaffe}), which means that the energy per baryon of
deconfined matter (a pure gas of quarks $u$ and $d$) at zero
pressure and temperature must be higher than $m_n$ (neglecting
refinements of order $\sim 1 \, MeV$ to the latter statement).
Work using the MIT-based EOS has shown that the latter condition
imposes that the MIT bag constant B must be larger than 57 MeV
fm$^{-3}$ (Farhi and Jaffe \cite{farhijaffe}).

Since all three flavors have the same common Fermi momentum, the absolute stability
condition takes a very simple form

\begin{equation}
\frac{\varepsilon}{ n_B} \bigg|_{P = 0} = 3 \mu \leq m_n = 939 MeV .
\end{equation}

\noindent Therefore, at the zero pressure point we have (without
any approximation)

\begin{equation}
B = - \Omega_{free}(m_s,\mu_0) + \frac{3}{\pi^2} \Delta^2 \mu_0^2 ,
\label{window}
\end{equation}

\noindent
where $\mu_0 = m_n /3  = 313$  MeV. A family of curves in the $m_s - B$ plane follow, and on each one
the energy per baryon is exactly $\varepsilon / n_B = m_n$ for a given $\Delta$ (Lugones and Horvath 2002).

Fig. 2 displays the stability window for the CFL phase (i.e. the
region in the $m_s$ versus B plane where $E/n_B$ is lower than
$939 \, MeV$ at zero pressure. Eq. (\ref{window}) sets the right
side boundary of the window while the left side boundary is
imposed by the minimum value B = 57  MeV. The window is
greatly enlarged for increasing values of $\Delta$ with respect to
the original SQM calculations which do not include any pairing
(see for example, Fig. 1 of Farhi and Jaffe \cite{farhijaffe}). We
have emphasized elsewhere that the pairing gap $\Delta$ may be important for a
"CFL strange matter" state (Lugones and Horvath \cite{cflprd}; see
also Madsen \cite{Madsen} for a similar view).

\section{Structure of color flavor locked stars}

\begin{figure}
\centering
\includegraphics[angle=-90,width=8cm,clip]{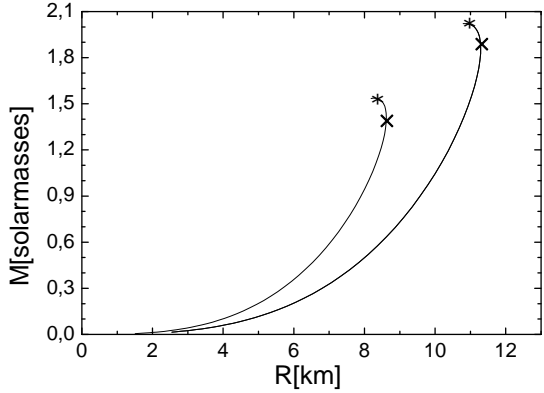}
\caption{The mass-radius relation for CFL strange stars for  B
=115 MeV fm$^{-3}$,  $m_s$ = 150 MeV, $\Delta =  100$ MeV on the
left, and for B = 70 MeV fm$^{-3}$,  $m_s$ = 150 MeV, $\Delta =
100$ MeV on the right. The crosses indicate the models of maximum
radius for each set of parameters, while the asterisks indicate
the models with the maximum mass (see Figs 4-7 for more details).}
\end{figure}

The study of the mass-radius relation for compact stars is a
widely known tool for testing the existence of different phases of
matter inside NSs. It is also well-known that simultaneous
observational data on masses and radii can impose important
constraints on the high density equations of state. Up to now,
almost all of the measured masses of "neutron stars" clustered
within a narrow range around 1.4 $M_{\odot}$. It has been
conjectured that this mass scale may be due to the fact that
neutron stars are formed in the gravitational collapse of
supernovae and come just from the iron cores of the
pre-supernovae. However, nothing fundamental precludes smaller
neutron stars from existing provided some mechanism capable of
creating them operates. Because small masses are not expected in
core collapse models, they hold the potential for discriminating
among possible equations of state. This is particularly
significant given the recent claims of very low-mass/low-radii
objects. As already mentioned, at least two sources, Her X-1 and
RX J1856-37, are candidates for high compactness. They have been
claimed to have radii $\sim 7 km$ and masses around $1 M_{\odot}$
(Li et al. \cite{li}; Pons et al. \cite{Pons}). Fragmentation of a
larger star (perhaps due to high rotation) may be a possible way
to produce sub-solar mass compact objects (Nakamura
\cite{tukulito}). Nevertheless, it is clear that models and data
are still subject to detailed analysis.

CFL strange matter offers the double bonus of enhancing the odds
for a self-bound state, and also of allowing the existence of
stable stars made up {\it entirely} of this phase. To pinpoint the
effects of such a composition we have solved the
Tolman-Oppenheimer-Volkoff equations of stellar structure,
assuming a parameter set inside the windows of absolute stability
discussed above (Lugones and Horvath \cite{cflprd}). Fig. 3 shows
sequences of compact star models thus calculated for particular
values of the parameters $\Delta$, $m_s$ and $B$ (see caption).
These values of the latter quantities have been selected to
highlight a potentially important feature of pairing in dense
quark matter, namely the possibility of obtaining very compact
stellar models. As expected, the mass-radius relation of stars
made up of CFL strange matter presents the same qualitative shape
as the curves found for strange stars . This is a direct
consequence of the existence of a zero pressure point at finite
density (of the order of the nuclear saturation density).

To assess the robustness of the results, and to understand the
outcome of masses and radii for a given set of parameters, we have
explored the effect of each parameters of the EOS on the static
structure of the stellar models. As shown in Figs. 4-7, and quite
analogously to the case of strange stars, the effect of increasing
$B$ while holding $\Delta$ and $m_s$ fixed is to lower both the
maximum radius $R_{max}$  and the maximum mass $M_{max}$ of the
stellar configurations. The mass of the strange quark $m_s$ works
in the same direction although the effect is much smaller. The
effect of the pairing gap $\Delta$ is qualitatively different and
much more interesting. An increase of $\Delta$ increases the value
of both $M_{max}$ and $R_{max}$. However, since the increase of
$\Delta$ strongly enlarges the stability window, the combination
of a large $\Delta$ with a large $B$ allows the construction of
very compact models made up of absolutely stable CFL strange
matter. This is particularly illustrative when we set $\Delta =
m_s / 2$. In this case the EOS adopts, to order $\mu^2$, the same
simple form $\varepsilon = 3 P + 4 B$ as unpaired SQM. However,
since the stability window is considerably enlarged, large values
of $B$ are allowed and very compact stars are found. This shows
how, although the effects in the EOS are only a few percent
($\cal{O}$ $(\Delta /\mu)^{2}$), the enlargement of the stability
window allows a very rich set of stellar configurations, ranging
from very compact to quite extended objects. As is seen from the
Figures, for the smallest values of $B$ the models tend to be less
compact and show radii up to 13 km and masses  as high as 2.4
$M_{\odot}$. The most extended configurations are obtained for a
combination of small $B$ and high $\Delta$. \footnote{{ It can be
checked that when the two particular values of B used by Alford
and Reddy (2002) are selected (B$^{1/4}$ = 165 MeV and B$^{1/4}$=
185 MeV; corresponding to B = 97 MeV fm$^{-3}$ and B = 153 MeV
fm$^{-3}$ respectively) the values obtained for $M_{max}$ and
$R_{max}$ from Figs. 4-7 are quantitatively in agreement (even if
the latter value has to be extrapolated if needed in these Figs.).
Moreover, a variety of CFL strange star models can be explored by
varying the parameters, although, for example, a large variation
in $m_{s}$ does not result in substantial modifications of
$M_{max}$ and $R_{max}$.}} It is not possible to select the most
likely range of the parameters, which should be limited by
observational arguments much in the same way as done for strange
matter models.

\begin{figure}
\centering
\includegraphics[angle=-90,width=8cm]{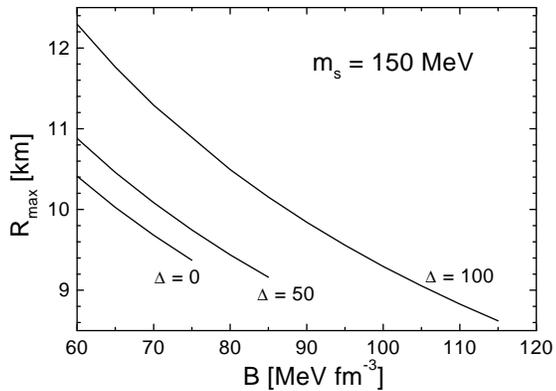}
\caption{The maximum radius indicated with crosses in  Fig. 3 is
shown here as a function of B, for   $m_s = 150$  MeV and
different values of  $\Delta$.}
\end{figure}

\begin{figure}
\centering
\includegraphics[angle=-90,width=8cm]{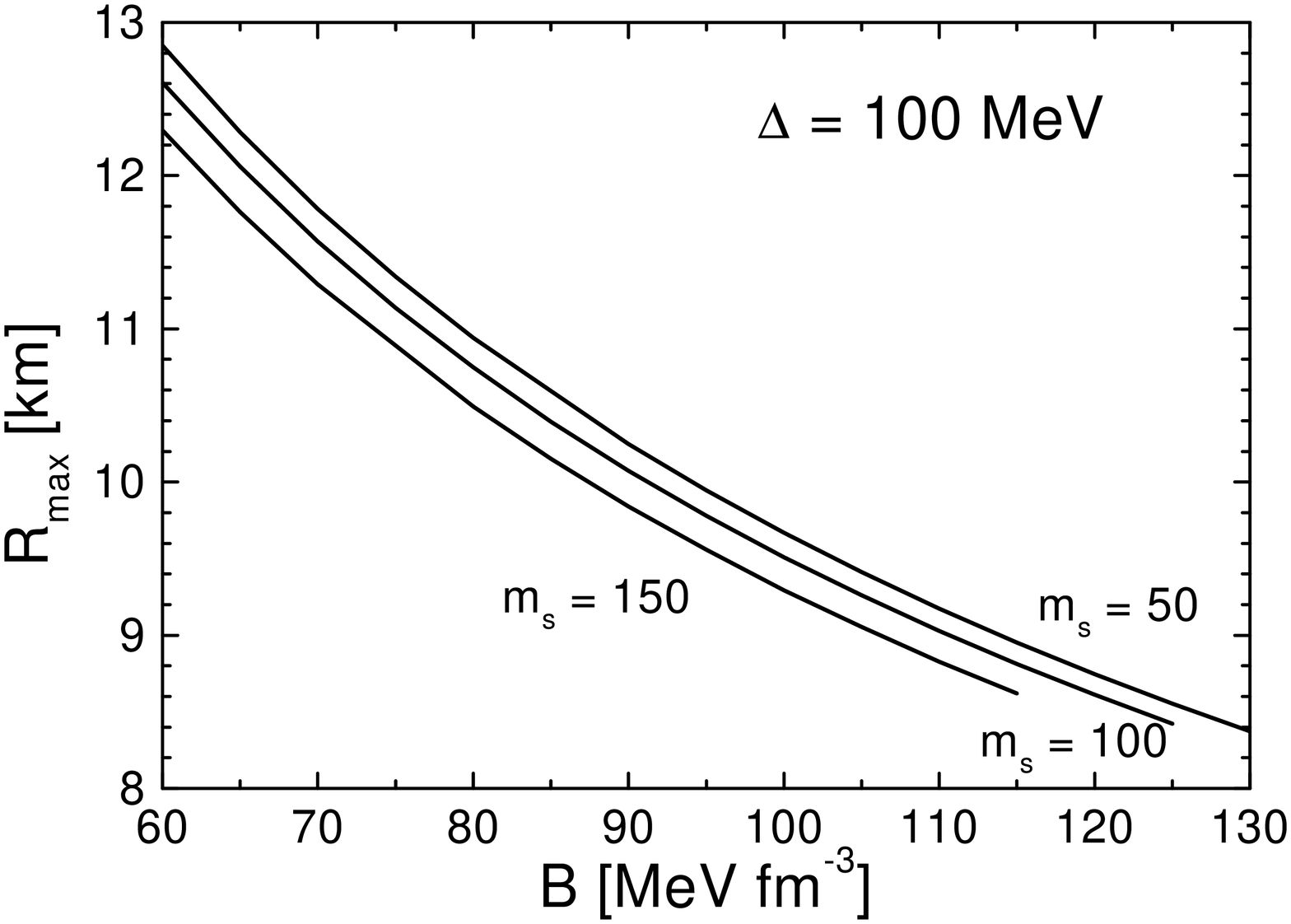}
\caption{The same as the previous figure but for fixed $\Delta$
and different values of $m_s$.}
\end{figure}

\begin{figure}
\centering
\includegraphics[angle=-90,width=8cm]{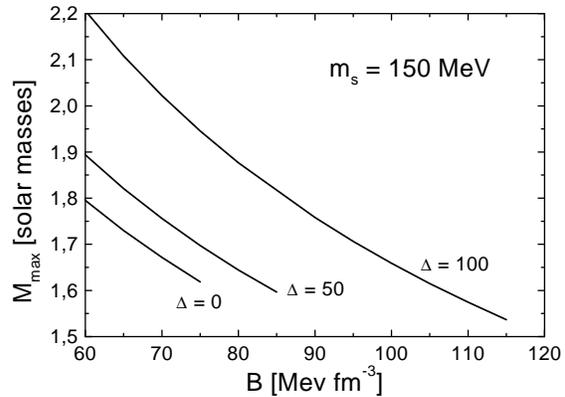}
\caption{The maximum mass indicated with asterisks in  Fig. 3 is
shown here as a function of B, for   $m_s = 150$  MeV and
different values of  $\Delta$.}
\end{figure}

\begin{figure}
\centering
\includegraphics[angle=-90,width=8cm]{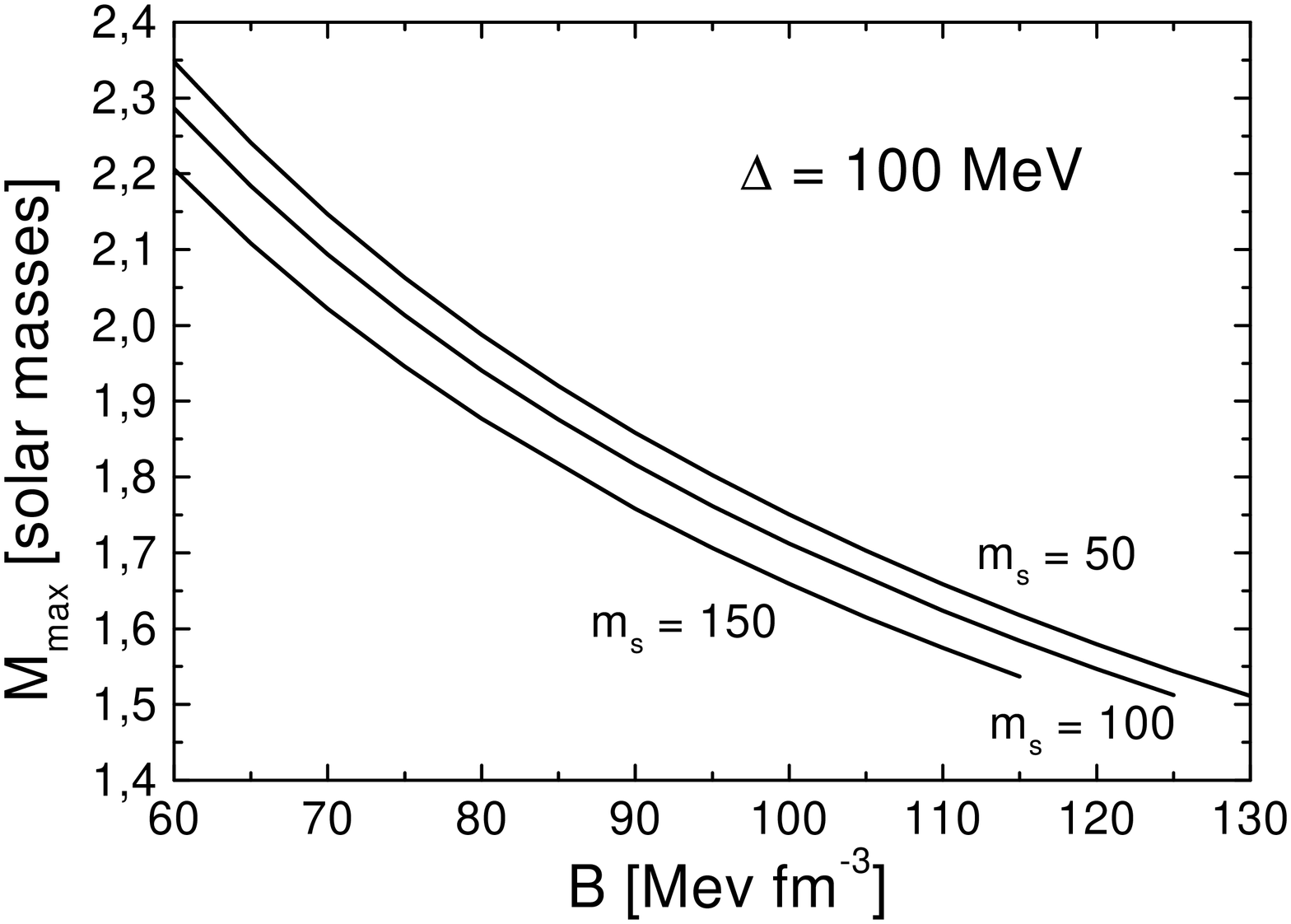}
\caption{The same as the previous figure but for fixed $\Delta$
and different values of $m_s$.}
\end{figure}

\section{Discussion}

We have addressed in this work stellar models constructed with the
simplest version of a "CFL strange matter" phase. The CFL phase at
zero temperature has been modelled as an electrically neutral and
colorless gas of quark Cooper pairs, in which quarks are paired in
such a way that all the flavors have the same Fermi momentum and
hence the same number density (Rajagopal and Wilczeck
\cite{rajagopal}). Since the strange quark is massive, some energy
must be spent in order to keep a common Fermi momentum for all
three flavors. However, more than that is gained from the energy
gap of the pairing. The model allows CFL strange matter to be the
true ground state of strong interactions for a wide range of the
parameters $B$, $m_s$ and $\Delta$. It is remarkable that the
strange matter hypothesis may be quite {\it boosted} by pairing
interactions. We have explored the mass-radius relation for all
the values of the parameter space of the EOS that give absolutely
stable CFL strange matter. Very compact configurations are found
that could help to explain the recently claimed compactness of a
few neutron stars. Also, for some parameter choices the models are
massive and extended.

A feature that has been largely explored and debated in the
context of strange star models is the existence of a normal matter
crust (Zdunik \cite{crust}). Bare quark surfaces may alter
drastically the ability of radiating photons, and thus they may
produce interesting signatures for their identification (Page and
Usov \cite{PU}). Conversely, a normal matter crust (held in
mechanical equilibrium by the electrostatic potential at the
surface) may hide most of the features of exotic matter. By its
very construction, it is clear that (in striking contrast with the
well-studied SS) no crust could be present in the case of CFL
strange stars. This is directly related to the absence of
electrons in the mixture (Rajagopal and Wilczek \cite{rajagopal}),
and thus to the absence of an electrostatic potential to prevent
normal matter from being converted to the stable CFL state. CFL
strange stars must have bare surfaces within these models. This in
turn means that the transport properties of the state, which are
still being explored (see, for example Shovkovy and Ellis
\cite{ISPE}; { Jaikumar, Prakash and Schafer 2002; Reddy,
Sadzikowski and Tachibana 2003}), hold the clue to the
identification of these compact stars. We expect the photon
emission properties of CFL strange stars to resemble those of bare
SS. Briefly, pairing effects should affect the plasma frequency
$\omega_p$ through the baryon number density as a small correction
of order $\mu \Delta^2$ so that $\omega_p$ will not be very
different from the $\sim 20 MeV$ expected for bare SS . The
equilibrium photon radiation will show a very hard spectrum and a
tiny luminosity, making CFL strange stars very difficult to
detect. On the other hand, while the thermal emission of photons
from the bare quark surface of a hot strange star (mainly by
electron-positron pair production) has been shown to be much
higher than the Eddington limit (Page and Usov \cite{PU}), this
would not be the case for CFL SS since no electrons will be
present at the surface.

{ As already mentioned, the fact of considering CFL SQM to be
absolutely stable together with the absence of electrons in the
mixture precludes the existence of a crust of normal matter, then
CFL strange stars are bare by construction. In a more general
approach, Alford and Reddy (2002b) found mass-radius relations
with essentially the same shape as that for neutron stars made up
of normal nuclear matter when considering a wider range of
parameters in which "hybrid" stars are allowed. This corresponds
to a large part of the parameter space and not only do they find a
normal matter envelope for all these models, but also the
composition of the less massive objects becomes pure nuclear
matter, as expected. In addition, as pointed out in Section 3, a
subset of the models of Alford and Reddy (2002) correspond to our
"CFL strange stars", and the full range of these self-bound models
can be obtained with the aid of Figs. 4-7.} The self-bound CFL
state may be compared to stable diquark states, which have been
found to produce very similar stellar structural properties than
CFL strange matter within completely different models (Horvath
\cite{SDM}; Horvath and de Freitas Pacheco \cite{HP}; Lugones and
Horvath \cite{vigilantes}). We finally note that, due to the
presence of $m_{s}$ and $\Delta$ in the EOS, a linear form
$P(\rho)$ is obtained in particular cases only, therefore the
scaling behavior of the Tolman-Oppenheimer-Volkov equations that
allows the construction of simple expressions for the maximum
masses and radii is lost for these CFL stars. The relevance of
this modelling to understand actual observed stars is not clear as
yet, but certainly the latter will continue to advance to the
point at which this and other questions will be answered with
confidence.

{\it Acknowledgements:} G. Lugones acknowledges the Instituto de
Astronomia, Geof\'\i sica e Ci\^encias Atmosf\'ericas de S\~ao
Paulo and the financial support received from the Funda\c c\~ao de
Amparo \`a Pesquisa do Estado de S\~ao Paulo. J.E. Horvath wishes
to acknowledge the CNPq Agency (Brazil) for partial financial
support.


\end{document}